\documentclass{nature}

\usepackage{lineno}
\usepackage{amsmath}
\usepackage{amssymb}
\usepackage{graphicx}
\usepackage{xcolor}
\usepackage{mathabx}
\usepackage{hyperref}

\makeatletter
\let\saved@includegraphics\includegraphics
\AtBeginDocument{\let\includegraphics\saved@includegraphics}
\renewenvironment*{figure}{\@float{figure}}{\end@float}
\makeatother

\title{Tidal fragmentation as the origin of 1I/2017 U1 (`Oumuamua)}

\author{Yun Zhang$^{1,2,3,4*}$ \&  Douglas N.C. Lin$^{5,2}$}

\begin{document}

\maketitle

\begin{affiliations}
 \item National Astronomical Observatories, Chinese Academy of Sciences, Beijing, China.
 \item Institute for Advanced Studies, Tsinghua University, Beijing, China.
 \item Universit\'e C\^ote d'Azur, Observatoire de la C\^ote d'Azur, CNRS, Laboratoire Lagrange, Nice, France.
 \item Department of Astronomy, University of Maryland, College Park, MD, USA.
 \item Department of Astronomy and Astrophysics, University of California, Santa Cruz, CA, USA.
 \item[*] To whom correspondence should be addressed; Email: \href{mailto:yun.zhang@oca.eu}{yun.zhang@oca.eu}
\end{affiliations}

\begin{abstract}
The first discovered interstellar object (ISO), `Oumuamua (1I/2017 U1)
shows a dry and rocky surface, an unusually elongated short-to-long axis 
ratio $c/a \lesssim 1/6$, a low velocity relative to the local standard 
of rest ($\sim 10$ km s$^{-1}$), non-gravitational accelerations, 
and tumbles on a few hours
timescale\cite{Bannister19, Meech17, Knight17, Bolin17, Fraser18, 
Drahus18, Micheli18, Trilling18, Jewitt17}. 
The inferred number density ($\sim 3.5 \times 10^{13} - 2 \times 10^{15}$
pc$^{-3}$) for a population of asteroidal ISOs\cite{Do18, Portegies18} 
outnumbers cometary ISOs\cite{Engelhardt17} by $\geq 10^3$, in 
contrast to the much lower ratio ($\lesssim 10^{-2}$) of rocky/icy 
Kuiper belt objects\cite{Weissman97}.  
Although some scenarios can cause the ejection of asteroidal 
ISOs\cite{Cuk18,Raymond18b}, a unified formation theory has yet to 
comprehensively link all `Oumuamua's puzzling characteristics and to 
account for the population.
Here we show by numerical simulations that `Oumuamua-like ISOs 
can be prolifically produced through extensive tidal 
fragmentation and ejected during close encounters of their 
volatile-rich parent bodies with their host stars. 
Material strength enhanced by the intensive heating during periastron 
passages enables the emergence of extremely elongated triaxial ISOs 
with shape $c/a \lesssim 1/10$, sizes $a \sim 100$ m, and rocky surfaces.  
Although volatiles with low sublimation temperature (such as CO) are 
concurrently depleted, H$_2$O buried under surfaces is preserved in 
these ISOs, providing an outgassing source without measurable cometary 
activities for `Oumuamua's non-gravitational accelerations during its 
passage through the inner Solar System.  
We infer that the progenitors of `Oumuamua-like ISOs may be km-sized 
long-period comets from Oort clouds, km-sized residual planetesimals 
from debris disks, or planet-size bodies at a few AU, orbiting around 
low-mass main-sequence stars or white dwarfs. These provide abundant 
reservoirs to account for `Oumuamua's occurrence rate.
\end{abstract}

We use the PKDGRAV $N$-body code to simulate 
the tidal disruption processes (see Methods).  
`Oumuamua's parent bodies are modelled as self-gravitating 
rubble piles, which represent the most typical structure 
for Solar System small bodies\cite{Richardson02, Mcneill18}.  
A soft-sphere discrete-element model is applied to compute 
the contact forces and torques between granules in the normal, 
tangential, rolling, and twisting directions.  These quantities 
determine the magnitude of the material shear and cohesive strengths.  
For the first set of models, the parent body is represented 
as a spherical granular aggregate with an initial bulk density 
of $\rho_p=2$,000 kg m$^{-3}$, a radius of $R_p=100$ m, 
and no initial spin. The host is assumed to
be a main-sequence star whose mass $M_s = 0.5 
M_{\odot}$, radius $R_s\simeq0.5R_\odot$, and 
average density $\rho_s \simeq 4 \rho_\odot$ 
(the region where tidal disruption occurs is much larger 
around low-mass stars as they have higher densities; see Methods). 
Its limiting distance of tidal disruption for gravity-dominated 
bodies\cite{Sridhar92} (such as gas giants, super-Earths, minor 
and dwarf planets) is $d_{\rm limit}(R_p\gtrsim1$ km$)
= 1.05(M_s/\rho_p)^{1/3}\simeq 8.3\times 10^8$ m $> R_s$. 
Small bodies with radii smaller than a few km are 
bonded by material strength. The limiting distance
for these strength-dominated bodies is modified to be 
$d_{\rm limit}(R_p\lesssim1$ km$) \simeq 
0.52(C_1/R_p^2 \rho_p^2 +C_2)^{-1/3}(M_s/\rho_p)^{1/3}$, 
where $C_1$ and $C_2$ are constants related to the material 
cohesive and shear strengths, respectively (see Methods). 
Objects with $d_{\rm limit} (R_p) \lesssim R_s$ would crash 
coherently into their host stars if their periastron distance 
$d_p<R_s$. Accordingly, our simulation results can be
scaled to different values of $R_p$, $\rho_p$, and $M_s$.

We performed a series of simulations with the parent 
body approaches the host star on a high-eccentricity
$e_0$ orbit with $d_p=3.5$--$8.0 \times 10^8$ m $> R_s$.  
After the tidal disruption, it breaks up into many pieces 
with fractional $f_E \sim \pm R_p/[(1-e_0) d_p]$ change to 
their incoming specific orbital energy.
The criterion for a significant fraction of 
the fragments to escape from the host stars 
is $\vert f_E \vert \gtrsim 1$ or $(1-e_0) 
\lesssim R_p/d_p$ (see Methods). For the parent bodies with
$R_p=100$ m considered in the first set of simulations,
$(1-e_0) \sim 10^{-6}$ is required 
to meet the ejection criterion. The corresponding initial
orbital semi-major axis $a_0=2$,300--5,300 AU,
which is comparable to the size of the inner Oort cloud 
in the Solar System.  The tidal debris of larger-$R_p$ parent 
bodies can readily escape with smaller $e_0$ and $a_0$.  
A second set of simulations (see below) on the passage of 
$R_p =2 \times 10^7$ m super-Earths with similar periastron 
distances shows that half of the fragments can be ejected 
with a smaller $e_0\sim 0.999$.  The corresponding $a_0$ 
(a few AU) is within the region where most planets reside.

\begin{figure}
\includegraphics[width=16.5cm]{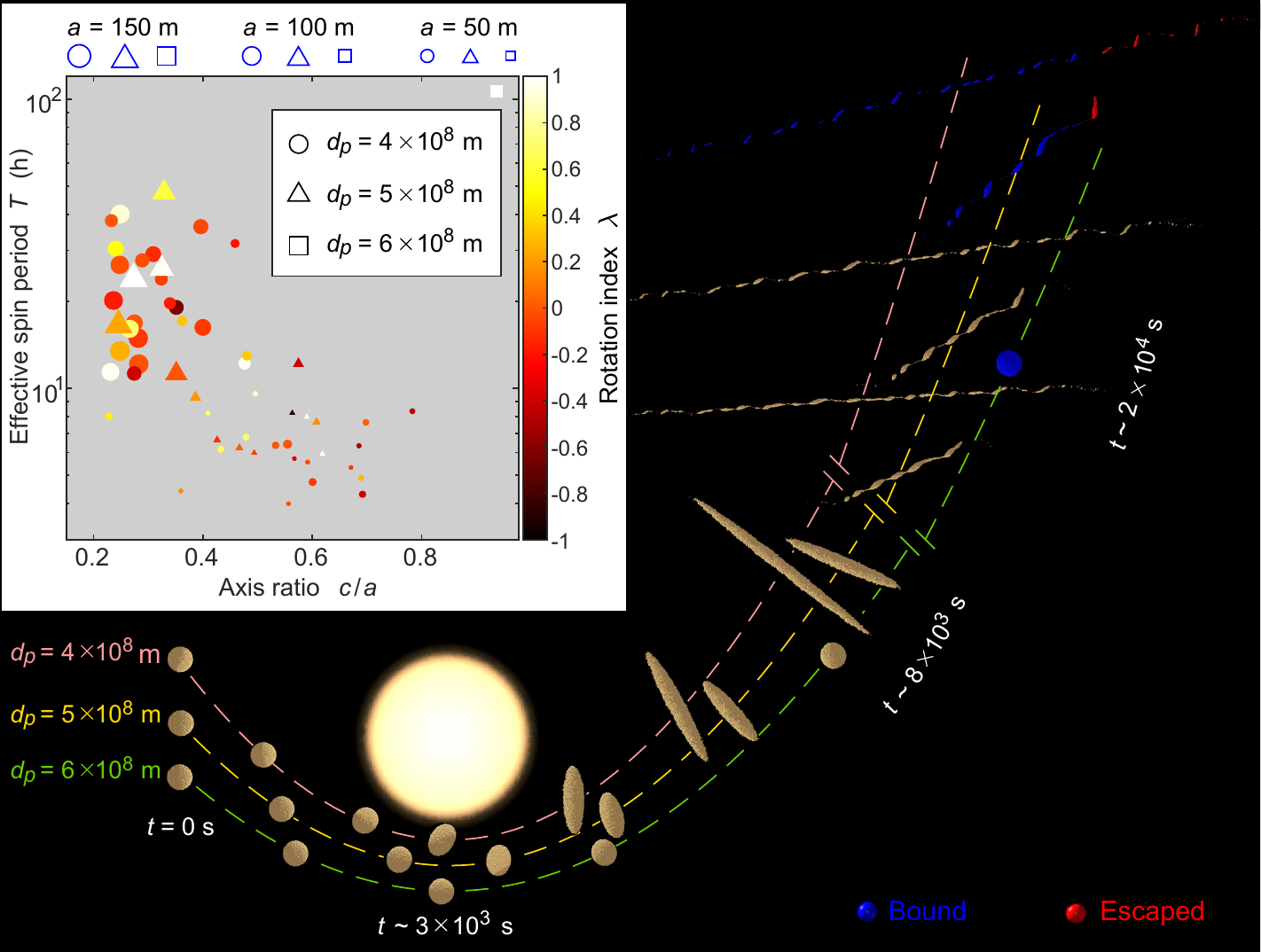} 
\caption{\textbf{Tidal disruption processes and fragmentation outcomes at 
different periastron distances for the first set of models.} The main panel 
superimposes the trajectories (the dashed lines) and the evolution of the 
rubble-pile parent body (the beige granular aggregates with mass centres 
locating at each corresponding trajectory; enlarged for illustration purpose) 
at different times $t$ as it flies by the star (from left to right), with 
three pericentre separations $d_p$. The luminous sphere represents the 
position and size of the star. The resulting fragments at the last scene are 
coloured by their orbit types, where the red ones can escape from their 
original planetary system and become interstellar objects. The top-left inset 
shows the distributions of the effective spin period $T$ and the short-to-long 
axis ratio $c/a$ of resulting fragments with mass $\ge10^6$ kg. The symbol 
size represents the semi-major axis $a$ for each fragment, and colour denotes 
the rotation index, where $\lambda=1$, 0, and $-1$ indicate the short-, 
intermediate-, and long-axis rotation states, respectively, and values in 
between indicate  non-principal-axis rotation states (see Methods). A friction 
angle of $35^\circ$ and a cohesive strength of 0 Pa are used in these 
simulations.}
\label{Figure1}
\end{figure}

As shown in Fig.~\ref{Figure1}, during the periastron passage, the 
integrity of the parent body is preserved by its intrinsic material 
shear strength for encounters with $d_p\gtrsim 6 \times 10^8$ m 
(see Methods). 
For closer encounters, the parent body is spun up, significantly 
distorted and then disrupted by the stellar tides, to produce 
numerous fragments (see Supplementary Video 1).  The tidal force 
is a steeply decreasing function of the periastron distance.  
With a smaller $d_p$, the parent body breaks up into a greater 
number of smaller, highly elongated fragments (see the top-left 
insert of Fig.~\ref{Figure1} and Supplementary Fig.~1).  
The normalised orbital-energy-per-mass increment $f_E > 1$ 
for some fragments. This change is adequate for them to 
escape (represented by the red fragments in Fig.~\ref{Figure1})
from the gravitational potential of their host star plus 
additional, if any, Jupiter-like planets (see Methods).

\begin{figure}
\includegraphics[width=16.5cm]{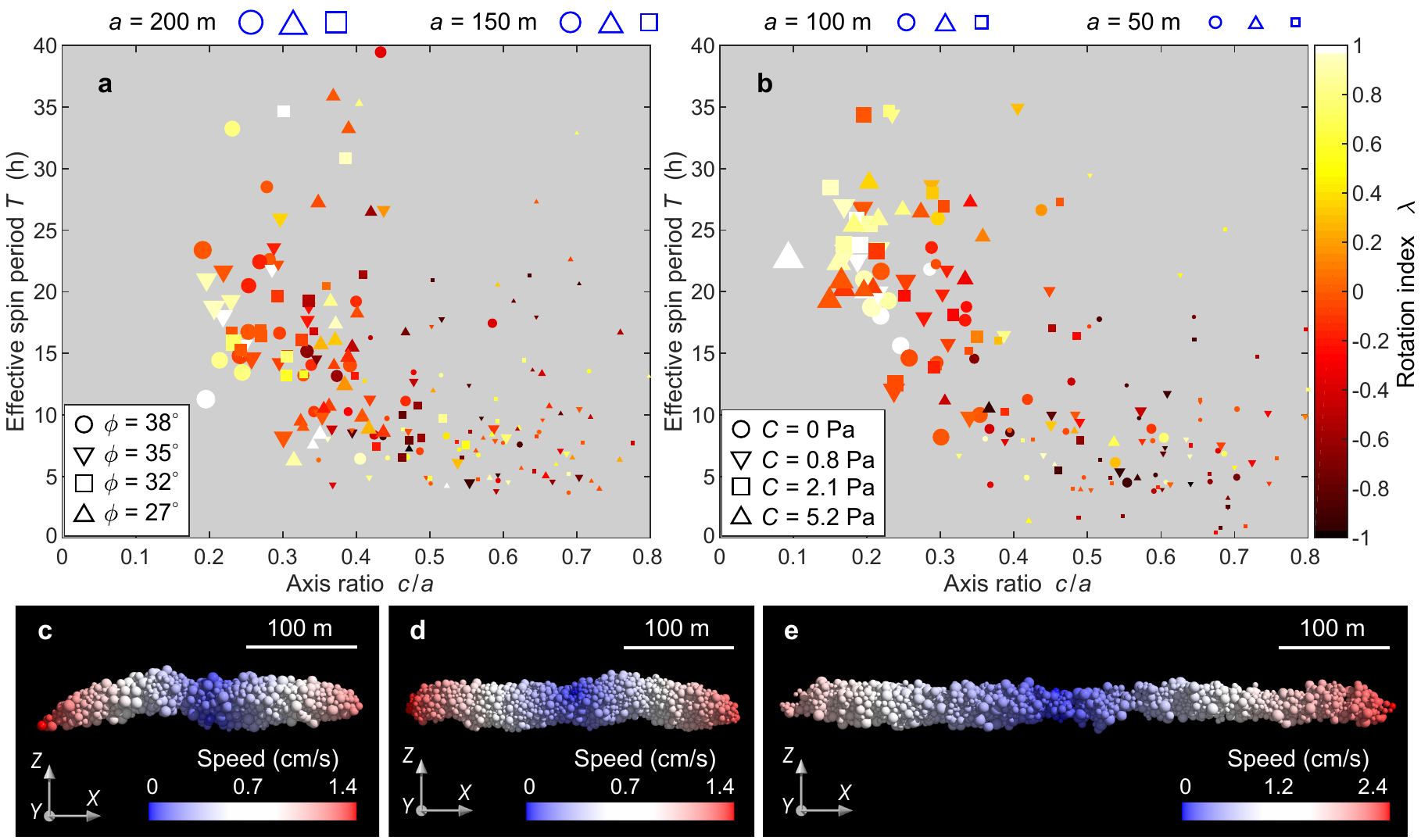} 
\caption{\textbf{Fragmentation outcomes of a range
of material strengths at $d_p = 3.5\times10^8$ m 
for the first set of models.} 
\textbf{a}, \textbf{b}, Effective spin period and axis ratio 
distributions of fragments with mass $\ge10^6$ kg for 
different material friction angles $\phi$ (\textbf{a}; $C=0$
Pa) and different later-turned-on material cohesive
strengths $C$ (\textbf{b}; $\phi = 35^\circ$). The
symbols have the same meanings as in the top-left
inset of Fig.~\ref{Figure1}. \textbf{c},
\textbf{d}, \textbf{e}, Examples of elongated
fragments formed by tidal disruption (\textbf{c, d}, 
$\phi = 35^\circ$, $C=0$ Pa; \textbf{e}, $\phi =
35^\circ$, $C=0$ Pa before crust formation at 3
hours post periastron and $C=5.2$ Pa after this transition),
where the colours show the bulk rotation.}
\label{Figure2}
\end{figure}

Nearly all the fragments are triaxial prolate and rapidly tumbling.
Larger fragments tend to tumble on timescale of a few to 
tens of hours with more extreme $c/a$ ratios.  
The light curve associated with this state is compatible 
with observations on `Oumuamua (see Methods).  
The timescale of damping their tumbling motion by their 
internal friction is $>$ 5 Gyr (see Methods). 
The fragments' internal density is generally reduced by the tidal 
and rotational deformation to $\sim \rho_p/2\sim 1$,000 kg m$^{-3}$. 
Despite the low density, their internal friction and cohesion 
(in addition to gravity) prevent them from rotational breakup. 
In order to consider a range of potential material strengths, we 
performed tidal disruption simulations with different friction angles 
$\phi$ and cohesive strengths $C$ to show that the production of 
elongated (with $c/a \lesssim 1/3$) fragments is robust 
(see Fig.~\ref{Figure2}).

\begin{figure}
\includegraphics[width=9cm]{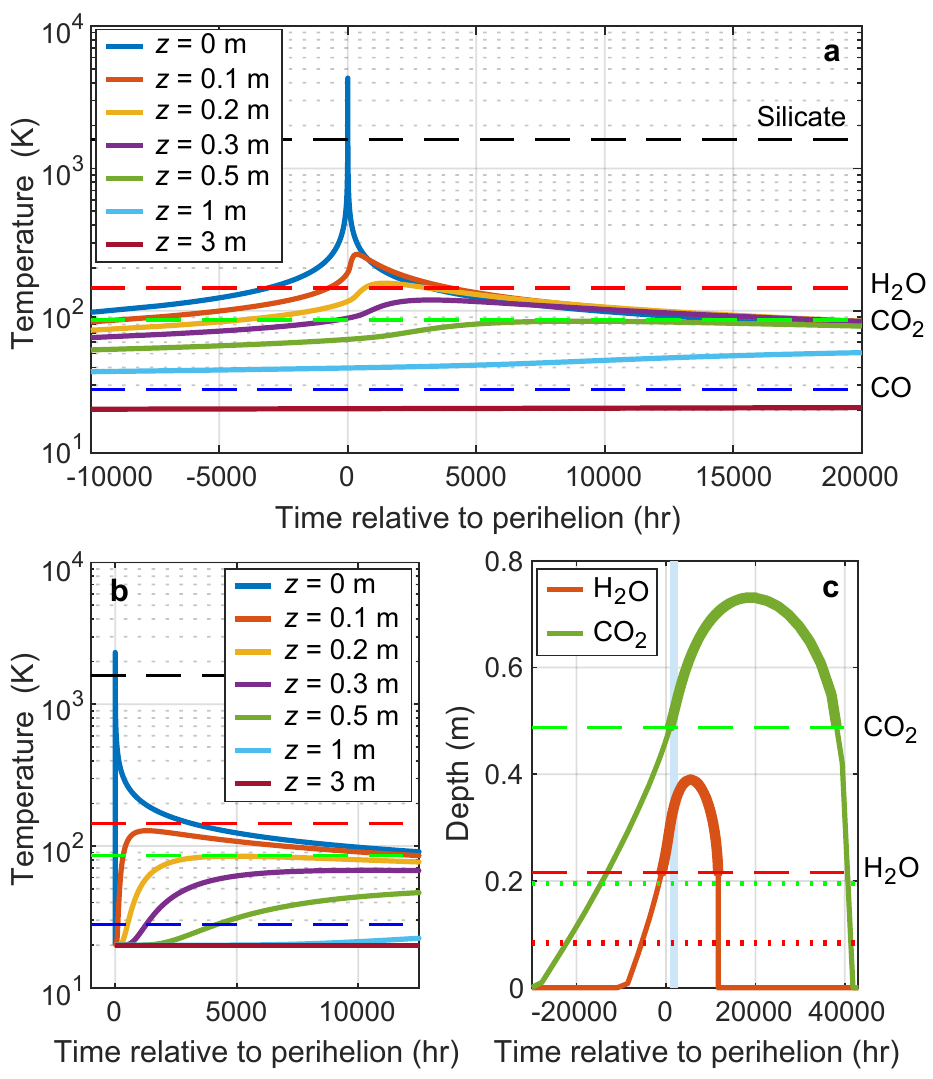} 
\caption{\textbf{Thermal modelling of a close
stellar flyby.} \textbf{a}, Temperature evolution
of the parent body during its tidal encounter with 
the host star ($M_s=0.5M_\odot$) on a parabolic orbit 
with a periastron distance $d_p = 3.5\times10^8$ m.
\textbf{b}, Temperature evolution of a fragment
with an initial temperature of 20 K, whose
surface is freshly exposed to the star after the
parent body is broken up during the periastron 
passage on the same orbit. In \textbf{a} and \textbf{b}, 
the temperature at various depths is represented by solid 
lines with different colours, and the vacuum sublimation
temperatures\cite{Gasc17} for H$_2$O, CO$_2$, and CO, 
and the melting temperature of silicate\cite{Lesher15} 
are represented by dashed lines. 
\textbf{c}, depth evolution of the H$_2$O and CO$_2$ 
sublimation layers during `Oumuamua's flyby of the Sun.
The red and green dashed/dotted lines represent the 
buried depths of residual H$_2$O and CO$_2$ ices in the 
formation scenario shown in \textbf{a}/\textbf{b}. 
The observation span of `Oumuamua is indicated by the 
light-blue region. Plenty of residual H$_2$O and CO$_2$ 
ices can commence sublimation during the Solar System 
passage as indicated by the bold lines.}
\label{Figure3}
\end{figure}

Planetesimal parent bodies originated well outside the 
snow line (at a few AU) are likely to have icy/rocky 
composition similar to that of the comets. But, their 
surfaces are intensely heated by their original host stars 
during the disruptive periastron passage.  The thermal 
modelling (see Fig.~\ref{Figure3} and Methods) shows that 
the melting and re-condensation (at 3 hours after periastron) 
of surface silicates lead to the build-up of desiccated 
crusts on the resulting fragments and the transformation 
from cometary to asteroidal exteriors. 
The enhancement of cohesion due to
sintering of silicates in the crusts leads to more extreme 
$c/a$ $(\lesssim 1/10)$ with $a \sim 100$ m 
(see Fig.~\ref{Figure2}b). These results
offer a possible formation scenario for the
reported shape, size, and surface features of `Oumuamua.

Heat diffusion continues after periastron passage.  
At 3 m beneath fragments' surfaces, the temperature may reach 
28 K (CO's sublimation temperature) at 37,000 days after 
periastron (at $\sim$100 AU), such that a substantial 
fraction of volatiles in an `Oumuamua-size fragment 
may be evaporated.  The CO depletion provides an explanation 
not only for `Oumuamua's spectroscopic properties\cite{Trilling18}, 
but also for the observationally inferred ``dryness'' (i.e., large 
rock/ice-ratio compared to cometary material) of the interstellar 
population\cite{Do18,Portegies18}.  Despite the CO depletion, 
H$_2$O and CO$_2$ (with higher sublimation temperature), 
buried at depths $>0.1$--0.2 m and $>0.2$--0.5 m, respectively, 
remain in a condensed form (see Fig.~\ref{Figure3}).

After the formation and ejection from their host stars, the 
fragments' motion is gravitationally accelerated, dragged
by the interstellar gas, and retarded by the interstellar 
magnetic field through Alfv\'en propulsion\cite{Drell65}.
When a dynamical equilibrium is established by the balance of 
these effects, the ISOs attain a size-dependent terminal speed 
$v_{\rm term}(R_\mathrm{ISO})$ relative to the local standard of 
rest (LSR) without significant spin evolution (see Methods).  
For a $100$-m-sized object formed a few Gyr ago, $v_{\rm term}$ 
is consistent with the observed motion of `Oumuamua 
($\sim$10 km s$^{-1}$) prior to its Solar System entry. 
This magnetic retardation effect is negligible on ISOs with radii 
$R_\mathrm{ISO} \gtrsim 1$ km and their motion relative to the LSR 
is expected to be similar to stars with comparable age.

Following `Oumuamua's passage through the inner Solar System, 
near perihelion, the H$_2$O sublimation temperature is reached 
at greater depths beneath the crusts of the fragments than during 
the course of their prior formation and ejection (see 
Fig.~\ref{Figure3}c), as the Sun is brighter and hotter 
than their original low-mass host star (see Methods). 
Vaporisation of this additional inventory of volatile may bring 
some organic material to their porous surfaces with photometric 
characteristics resembling that found from `Oumuamua's colour 
analyses\cite{Fitzsimmons18}.  
Despite the lack of coma, the delayed resumption of outgassing 
may also lead to the reported non-gravitational 
accelerations\cite{Micheli18, Trilling18} (see Methods).

Based on the described formation and ejection mechanism, 
long-period comets (LPCs) and residual planetesimals 
on nearly parabolic orbits are promising candidates 
for `Oumuamua's parent bodies. Under the combined influence 
of the Galactic tide, passing stars, and planetary 
perturbations\cite{Kaib09, Punzo14}, a few km-sized LPCs 
intrude each year from the inner Oort cloud 
(with $a_0 \lesssim 2 \times 10^4$ AU)
to the stellar proximity with $1-e_0 \lesssim 10^{-5}$.
Since the size of typical visible comets are much larger 
($ \gtrsim$10) than `Oumuamua, they can provide 
an adequate supply of asteroidal ISOs through the accumulative 
ejection of the extensively downsized tidal fragments 
(analogous to Shoemaker-Levy-9) from a few percent of the LPC 
population (comparable to the required mass ejection per star 
$M_{\rm ej}\simeq(0.0017$--$0.1)M_\oplus$; see Methods).  
Debris disks are common around other stars and they are 
also potential reservoirs of asteroidal ISOs. 
Similar physical processes can also lead to cometary 
ISOs\cite{Portegies18} through the direct detachment of icy 
residual planetesimals without tidal disruption and intense 
heating in the proximity of their host stars (which are the 
critical processes to produce attributes similar to `Oumuamua).

Kepler\cite{Coughlin16} and microlensing\cite{Cassan12} 
surveys unveiled the omnipresence of multiple close-in 
super-Earths and distant sub-Neptunes around low-mass stars. 
These planets provide a second class of plausible
parent-body candidates.  Since a planet's 
mass $M_p \gg M_{\rm ej}$, a few percent efficiency 
per star is adequate to account for the asteroidal 
ISOs' occurrence rate. In multiple systems, 
their eccentricity can be elevated to unity by their own 
dynamical instability\cite{Nagasawa08, Idaetal13}.  
Our simulations of the tidal disruption of a 
self-gravity-dominated super-Earth (with $R_p=2 \times 10^7$ m 
and $M_p=11 M_\oplus$) on an $e_0 =0.999$ and $d_p=4 \times 
10^8$ m orbit around a $0.5 M_\odot$
host star show that half of the fragments, with mass
up to $M_{\rm upper} \sim 0.1 M_p$, can obtain a normalised 
orbital-energy-per-mass increment $f_E > 1$ 
and escape from the host star (see Methods).  
Through extremely close encounters with $d_p \ll d_\mathrm{
limit}$, super-Earths can be reduced to fragments with 
$M_{\rm upper} \ll M_p$. Tidal downsizing continues until 
the fragments are small enough to be dominated by material 
strength without further disintegration (see Methods).  
Since for the host stars $R_s \propto M_s$ and 
$d_\mathrm{limit}/R_s \propto M_s^{-2/3}$, tidal fragments with 
$a \sim 100$ m are produced more readily in the proximity 
of low-mass main-sequence stars.
These hosts are also the most common and long-lasting stars 
with ubiquitous multiple super-Earths\cite{Charbonneau09}. 
A modest planet/star mass ratio is favourable for the planets 
to acquire large $e_0$ through dynamical instability and secular 
chaos\cite{Davies14}, and for their fragments to be ejected. 
During the cooling and solidification of tidally 
disrupted fragments of planets' molten cores, sintering
surfaces provide adequate material strength to produce
the observed $c/a$ ratio.  However, the impact of their 
thermal evolution on the retention of volatiles to account for 
`Oumuamua's non-gravitational accelerations is highly 
uncertain (see Methods).

Similar physical processes around white dwarfs offer a third 
pathway for the prolific production of tidal 
debris\cite{Raymond18b, Rafikov18}, with a fraction of which 
may have rocky surfaces and sizes similar to `Oumuamua.
As the post-main-sequence byproducts of stars,
white dwarfs may retain some residual icy 
planetesimals and planets after their progenitors have lost 
their red-giant envelopes\cite{Veras11}. 
The total population of their progenitors is largely outnumbered 
by that of main-sequence stars with $M_s \lesssim 0.5M_\odot$. 
They may contribute a small fraction of the original hosts
of asteroidal ISOs (see Methods).

In our attempt to address plausible causes
of all aspects of the `Oumuamua conundrum, we
highlight the prolificacy and robustness of asteroidal
ISO diffusion between stars near and far. Since these sojourns 
pass through the domains of habitable zones, the prospect of 
panspermia, carried by them (nicknamed 
{\it sola lapis}\cite{Portegies18}), cannot be ruled out.


\begin{addendum}
\item[Correspondence and requests for materials] 
Correspondence and requests for materials should be 
addressed to Y.Z.
\item[Acknowledgements]  Y. Z. acknowledges funding from the Universit\'e 
C\^ote d'Azur 
``Individual grants for young researchers'' program of 
IDEX JEDI. 
D.N.C.L. thanks Institute for Advanced Studies, Princeton
for support while this work was initiated. We thank Scott
Tremaine for inspiration and suggestions, Derek 
C. Richardson for assistance with the PKDGRAV code, 
Gregory Laughlin, Patrick Michel, Shangfei Liu, Roman
Rafikov, and Simon Portegies Zwart for constructive 
feedback on the results and implications of this work. 
Simulations were carried out at the University of 
Maryland on the yorp cluster administered by the 
Department of Astronomy and the deepthought and 
deepthought2 supercomputing clusters administered by the 
Division of Informational Technology. For data 
visualisation, the authors made use of the freeware, 
multiplatform, ray-tracing package, Persistence of 
Vision Raytracer.
\item[Author contributions] Y.Z. performed the 
soft-sphere/$N$-body numerical simulations, the thermal 
modelling, and analysed the numerical results and 
implications for `Oumuamua. D.N.C.L. initiated the 
collaboration to study tidal disruption as a formation 
mechanism for `Oumuamua, and contributed to address questions 
on ISOs' population and dynamical origins. 
Both authors contributed to interpretation of `Oumuamua's
properties and preparation of the manuscript. 
\item[Author information]  
\textit{National Astronomical Observatory of China, Beijing, China}
Yun Zhang

\textit{Institute for Advanced Studies, Tsinghua University, Beijing, China}
Yun Zhang \& Douglas N. C. Lin

\textit{Universit\'e C\^ote d'Azur, Observatoire de la C\^ote d'Azur, CNRS, Laboratoire Lagrange, Nice, France}
Yun Zhang

\textit{Department of Astronomy, University of Maryland, College Park, MD, USA}
Yun Zhang

\textit{Department of Astronomy and Astrophysics, University of California, Santa Cruz, CA, USA}
Douglas N. C. Lin
\item[Competing interests] The authors declare no 
competing interests.
\end{addendum}

\newpage
\begin{methods}

\subsection{Soft-sphere discrete element model.}
We model the parent body as a self-gravitational aggregate 
consisting of $\sim$20,000 
spheres with a $-3$-index power-law particle size distribution (i.e., 
$\mathrm{d}\ln{N}/\mathrm{d}\ln{S}=-3$, where $N$ is the number of 
particles and $S$ is the particle size).  
Within the high-efficiency parallel 
tree code framework, PKDGRAV\cite{Richardson00,
Stadel01}, a soft-sphere contact model is used
for computing particle contact forces with four
components in the normal, tangential, rolling, 
and twisting directions\cite{Schwartz12, Zhang17,
Zhang18}. The compressive strength of the
material is controlled by two stiffness
constants, ($k_N$, $k_S$), for the normal and
tangential directions; the contact energy
dissipation is controlled by two coefficients of
restitution, ($\varepsilon_N$, $\varepsilon_S$),
for the normal and tangential directions; the material 
shear strength is controlled by three friction coefficients for 
the tangential, rolling, and twisting directions, 
($\mu_S$, $\mu_R$, $\mu_T$), 
and a shape parameter, $\beta$; and the material cohesive strength 
is controlled by an interparticle cohesive tensile strength, 
$c_p$, along the normal direction. 

The contact model as well as the relation between parameter setup and 
material strengths have been calibrated with laboratory experiments on 
real sands\cite{Schwartz12, Zhang18}. To precisely integrate the 
interactions between particles, the timestep $\Delta t$ is set to 
0.01 s for the simulations considered here. 
The normal stiffness $k_N$ is set to ensure that particle overlaps do not 
exceed 1\% of the minimum particle radius. The tangential stiffness $k_S$ 
is set to $(2/7)k_N$ to keep normal and tangential oscillation frequencies 
equal to each other. The coefficients of restitution,  $\varepsilon_N$ and 
$\varepsilon_S$, are set to 0.55 resembling the energy dissipation behaviours 
of terrestrial rocks\cite{Chau02}. $\mu_R$ and $\mu_T$ are set to 1.05 and 
1.3, respectively, corresponding to sand particles of medium 
hardness\cite{Jiang15}. The free parameters $\mu_S$ and $\beta$ are used 
to adjust the material friction angle\cite{Zhang17} $\phi$, for which
$\mu_S=0.2$ and $\beta=0.3$ corresponds to a friction angle of 27$^\circ$,
$\mu_S=0.5$ and $\beta=0.4$ corresponds to a friction angle of 32$^\circ$,
$\mu_S=1.0$ and $\beta=0.5$ corresponds to a friction angle of 35$^\circ$,
and $\mu_S=1.3$ and $\beta=0.7$ corresponds to a friction angle of 
38$^\circ$. The interparticle cohesion $c_p$ is used\cite{Zhang18} to
adjust the material cohesive strength $C$, 
for which 
$c_p=100$ Pa corresponds to $C=0.8$ Pa, $c_p=200$ Pa corresponds to 
$C=2.1$ Pa, and $c_p=400$ Pa corresponds to $C=5.2$ Pa for a constant 
friction angle of 35$^\circ$.

Fig.~\ref{Figure2}a shows more 
elongated fragments are formed with a larger friction angle.  
This dependence is expected because higher shear resistance 
can slow down the relative motion between the constituent particles 
as well as maintain the stability of more extreme shape.  
Normally, the friction angle of granular material 
is in the range of 25$^\circ$ to 45$^\circ$.  Our numerical experiments 
show that extremely elongated fragments with axis ratios of $\sim$0.3 
can form even for a material friction angle as low as 27$^\circ$. 
This implies that production of elongated fragments through 
tidal disruption is robust.  Nevertheless, 
the particles' relative motion is hard to be damped out completely by 
friction alone and $c/a < 0.2$ is hard to achieve even using a 
friction angle above 40$^\circ$.  The inclusion of initial,  
constant cohesion into the parent body's material also cannot help 
to form a more elongated fragment (see Supplementary 
Fig.~2 and Supplementary Video 2). 

However, the material cohesive strength of the parent body and its 
fragments can be changed by the thermal effect during their periastron 
passage.  In proximity of their original host, 
their exposed surfaces could be briefly heated by the intense stellar 
radiation to a temperature above the melting point of silicates, 
i.e., $T \geq1600$ K (see Fig.~\ref{Figure3}).  
On their outward journey, the solidification 
of their surfaces facilitates the formation of sintering bonds between 
surface particles\cite{Schwartz13,Poppe03}.  The enhanced cohesive strength 
inhibits subsequent break-apart and enables the formation and survival 
of some very elongated fragments.  We carried out a series of 
phase-transition simulations in which cohesion between particles in the 
parent body is imposed 3 hours after the periastron passage 
(when the silicates can re-condense).  
The production of fragments with $c/a\lesssim0.1$ under such a 
circumstance is shown in Fig.~\ref{Figure2}b and Supplementary Video 3.

We also explored the effects of shape, initial rotation state, and 
particle size distribution and resolution of the parent body in the 
tidal disruption simulations (see Supplementary Figs.~3--5). 
The production of elongated fragments with $c/a < 0.3$ is robust 
in all the simulations.

\subsection{Tidal disruption limiting distance.}
When an object approaches a star within the Roche limit, 
tidal forces imposed by the star overwhelm the gravitational 
forces that hold the object together and may disrupt it.  
The Roche limit of an initially spherical,
non-spinning, viscous fluid object is\cite{Sridhar92}

$d_\mathrm{Roche}=1.05\Big(\dfrac{M_s}{\rho_p}\Big)^{1/3}$, 

\noindent where $\rho_p$ is the bulk density of the object and $M_s$ 
is the mass of the star. 
$d_\mathrm{Roche}$ is appropriate to predict the tidal disruption 
limiting distance for bodies in the gravity-dominated regime, 
such as gas giants, super-Earths with molten interiors, minor 
and dwarf planets.

While for small bodies in the material-strength-dominated
regime, their intrinsic material shear and cohesive strengths 
can prevent tidal disruption within this limit.
To reflect the material characteristics, we use 
the elastic-plastic continuum theory\cite{Holsapple08} 
to estimate the tidal disruption limit of small bodies in 
the material-strength-dominated regime,

$d_\mathrm{str} = \Big(\dfrac{\sqrt{3}}{4\pi}\Big)^{1/3}\Big(\dfrac{5C}{4\pi GR_p^2\rho_p^2}+\dfrac{2\sin{\phi}}{\sqrt{3}(3-\sin{\phi})}\Big)^{-1/3} \Big(\dfrac{M_s}{\rho_p}\Big)^{1/3}$,

\noindent which depends on the radius of the object $R_p$, 
and decreases with either a larger friction angle $\phi$ 
or a stronger cohesive strength $C$. 
$G$ is the gravitational constant. 
This limit is much smaller than the Roche limit, e.g., 
even for low-strength materials with $\phi=25^\circ$ and $C = 0$ Pa, 
$d_\mathrm{str} = 0.9 (M_s/\rho_p)^{1/3} < d_\mathrm{Roche}$.
Material strength in small bodies provides a stabilising 
effect against the host star's tidal perturbations.

When the object's cohesive strength $C>0$
Pa, $d_\mathrm{str}$ is size-dependent. 
For example, during a close flyby with a star with 
$M_s =0.5M_{\odot}$ and $R_s=0.5 R_\odot\sim 3.5 \times 10^8$ m, 
a cohesionless object with $\rho_p=2$,000 kg m$^{-3}$ 
and $\phi=35^\circ$ can be tidally distorted if the periastron 
distance $d_p \le d_\mathrm{str} = 6.3 \times 10^8$ m. 
If its material contains a cohesive strength $C=5.2$ Pa, 
and its radius $R_p=100$ m, $d_\mathrm{str}$ would be decreased 
to $4 \times 10^8$ m.  
For a 10-m-radius object with equivalent material 
strengths,  $d_\mathrm{str}\simeq 9.5\times10^7$ m $<R_s$, 
indicating that smaller bodies are much harder to be 
tidally disrupted outside the stellar surface.  
It also indicates that $100$-m-sized fragments with some cohesion
can be preserved without further fragmentation outside $R_s$.
These analyses are consistent with our simulation results, 
which show that an $100$-m-radius object with equivalent 
properties and no cohesion is slightly distorted when passing 
by the star at a distance of $6\times 10^8$ m (see 
Fig.~\ref{Figure1}) whereas it can preserve its integrity with 
$C=5.2$ Pa at a smaller $d_p$ ($\sim 4 \times 10^8$ m).  

As the transition between regimes occurs at a few km radius, 
the limiting distance for tidal disruption is given by 
$d_\mathrm{Roche}$ in the gravity-dominated regime and
$d_\mathrm{str}$ in the material-strength-dominated regime as

$d_\mathrm{limit}(R_p)=\begin{cases}
d_\mathrm{Roche},\quad R_p \gtrsim~1~\mathrm{km} ; \\
d_\mathrm{str},\quad R_p \lesssim~1~\mathrm{km}. 
\end{cases}$ 

\noindent Since $d_\mathrm{limit}\propto(M_s/\rho_p)^{1/3}$, 
our simulation results can be scaled to different stellar 
mass $M_s$ and parent-body bulk density $\rho_p$.

As $R_s \propto M_s$ for main-sequence stars, 
$d_\mathrm{limit}/R_s \propto M_s^{-2/3}$, 
the region where tidal disruption occurs,
i.e., $R_s< d_p \lesssim d_\mathrm{limit}$,
is much larger around lower-mass stars. 
For most of the Solar System bodies, 
$d_{\rm limit} \lesssim R_\odot$, in which case they would
crush into the Sun's surface before tidal fragmentation.  
Some solar twin binary stars with planets exhibit signs of 
super-Earth ingestion\cite{Ramirez15}. 
This fragmentation barrier also accounts for the absence of 
small bodies in the Solar System with an extreme shape 
like `Oumuamua.  
In contrast, $d_\mathrm{limit}(R_p\gtrsim 1~\mathrm{km})$
is much larger than $R_s$ for stars with $M_s < 0.8 M_\odot$. 
For a moderate cohesive strength $C = 10$ Pa, which can be 
readily provided by the sintering bonds formed during close 
stellar encounters, the radii of low-mass stars 
are comparable to $d_\mathrm{limit}(R_p\sim 10$--100 m$)$.
Therefore, planets and planetesimals around low-mass stars
can be efficiently tidal-downsized to sub-km-sized fragments. 

Most recently emerged white dwarfs in globular star 
clusters\cite{Kalirai09} have $M_s \ge 0.53 M_\odot$
and $R_s \sim 10^{-2}R_\odot$. 
Older isolated white dwarfs have comparable $R_s$ and $M_s$.  
Since their $d_\mathrm{limit} \gg R_s$, 
there is a vast range of $d_p$ for tidal downsizing of 
planets and residual planetesimals.
For a moderate cohesive strength $C = 10$ Pa, this 
downsizing can powderise these bodies into centimetre-sized 
fragments at a periastron distance $d_p\sim R_s$. 
The production and preservation of `Oumuamua-size ISOs 
(with $a\sim 100$ m) requires the parent body passing 
the host at a distance of 
$d_p\sim d_{\rm limit} (100 {\rm~m}) \sim 0.5 R_\odot$.

\subsection{Ejection condition.}
As a single entity, the parent body passes through its periastron with
an orbital speed $V_p = \sqrt{ (1+e_0) G M_s/d_p}$ in the direction 
${\bf\hat {V}}_p$, a unit orbital angular velocity vector 
${\bf \hat{\Omega}} = {\bf\hat d}_p \times { {\bf\hat V}_p}$ and 
total energy per unit mass  

$E_0 = - \dfrac{G M_s (1-e_0)} {2 d_p}.$

\noindent
Under the intense tidal torque, the parent body acquires a prograde 
uniform spin angular frequency $\Omega_f = f V_p/d_p$, 
where $f$ is a factor of order unity. 
After its tidal disruption, the fragments at distance $r_f$ from 
the parent body's centre of mass (in the direction ${\bf \hat{r}}_f$) 
acquire a modified velocity (relative to the host star) 

${\bf V}_p^\prime = {\bf V}_p +
\Omega_f r_f { {\bf\hat{\Omega}} \times {\bf\hat {r}}_f}.$
  
\noindent
For simplicity, we consider those fragments at $r_f = \pm R_p/2$ (i.e., 
half of the parent body's size on the far and near side to the host 
star, relative to the parent body's centre of mass) and set 
${\bf\hat {r}}_f={\bf\hat {d}}_p$.  Since the individual 
fragments are no longer bound to each other, their energy per unit 
mass is modified to 

$E_{\pm} ^\prime = -\dfrac{GM_s}{d_p \pm R_p/2} + \dfrac{({V}_p \pm
\Omega_f R_p/2)^2}{2} \simeq -\dfrac{G M_s (1-e_0)}{2 d_p} 
\pm \dfrac{G M_s}{2 d_p} \dfrac{(1+2f) R_p }{d_p}.$

\noindent
After the tidal disruption of the parent body, the fractional incremental energy change  

$\vert f_E \vert = \dfrac{\vert E_{\pm} ^\prime - E_0\vert}{ \vert E_0 \vert} 
\simeq \dfrac{(1+2f)}{(1-e_0)} \dfrac{R_p}{d_p}$ 

\noindent
of the fragments in its outer region (with $r_f > R_p/2$) exceeds unity for 

$1-e_0 \lesssim 1-e_{0,\mathrm{crit}}=(1+2f) \dfrac{R_p}{d_p}$. 

\noindent
In this limit, up to half of the fragments acquire sufficient energy to 
escape from the gravitational confinement of the host star. Since 
$(1-e_{0,\mathrm{crit}}) \propto R_p$, this ejection criterion is more 
stringent for small-size parent bodies.

Around host stars that bear planets (with 
mass $M_J$ and orbital semi-major axes $a_J$), additional
gravitational binding energy $\Delta E = - G M_J
/ d$ needs to be taken into account at distance
$d > a_J$.  Since the post-periastron energy
$E^\prime$ is conserved at $d < a_J$, the 
criterion for fragments to escape
is $E^\prime > \Delta E$ or equivalently, 
 
$1-e_0 \lesssim \dfrac{(1+2f)}{(1+2 M_J a_0/M_s a_J)} \dfrac{R_p}{d_p}.$

\noindent 
The modification due to the Jovian-mass planet is
generally small unless the initial $a_0 \gg a_J$.
Parent bodies would be retained in the regions
beyond the planets (analogous to the Kuiper
belt and the Oort cloud in the Solar System), if
their

$ R_p \lesssim \dfrac{2}{(1 + 2f)}\dfrac{M_J}{M_s}\dfrac{d_p^2}{a_J}.$

\subsection{Rotation state.}
The rotation state of a body can be characterised by the effective spin 
period $T$ and the dynamic inertia $I_D$, which are the spin period and moment 
of inertia of a sphere with an equivalent total rotational energy and momentum, 
respectively\cite{Scheeres00}. The rotation mode of the body can be inferred 
from the value of $I_D$. With principal moments of inertia $I_z>I_y>I_x$, the 
body rotates along its maximum moment of inertia if $I_D=I_z$; the body's spin 
axis precesses along its maximum moment of inertia (i.e., short-axis mode, SAM) 
if $I_z>I_D>I_y$;  the body rotates along its intermediate moment of inertia if 
$I_D=I_y$; the body's spin axis precesses along its minimum moment of inertia 
(i.e., long-axis mode, LAM) if $I_y>I_D>I_x$; and the body rotates along its 
minimum moment of inertia if $I_D=I_x$.  Accordingly, we define the rotation 
index $\lambda$ as

$\lambda=\begin{cases}
(I_D-I_y)/(I_z-I_y),\quad I_D\ge I_y ; \\
(I_D-I_y)/(I_y-I_x),\quad \mathrm{otherwise}. 
\end{cases}$ 

\noindent Therefore, $\lambda=1$, 0, $-1$ indicates the principal-axis 
rotation state about the maximum, intermediate, and minimum moment of 
inertia, respectively. $0<\lambda<1$ indicates SAM and $-1<\lambda<0$ 
indicates LAM.

Our simulations show that the fragments of tidal disruption primarily 
have prolate triaxial shape (see Figs.~\ref{Figure1}, \ref{Figure2}, 
and Supplementary Fig.~1). 
Although brightness variations of `Oumuamua indicate an extremely 
small short-to-long axis ratio, the shape of this object is poorly 
constrained due to lack of knowledge about its rotation state and 
albedo variation\cite{Belton18}.  
The most comprehensive model published to date\cite{Mashchenko19} found 
that the large-amplitude variation in `Oumuamua's light curve can be 
interpreted as a tumbling prolate object with an axis ratio of 
$c/a \sim$1/8 or a tumbling oblate object with an axis ratio of 
$c/a \sim$1/6.  The probabilities of reproducing light curve minima as 
deep as the observed values with a randomly oriented angular momentum 
vector are 16\% for the best-fit 1/8 prolate and 91\% for the 1/6 
oblate model. Since the value of the axis ratio depends on the orientation 
of the rotation pole, a more comprehensive probability comparison is 
warranted for a range of prolate and oblate triaxial models (in addition 
to the best-fit models).

Among all of the small bodies in the Solar System with 
high-quality shape data\cite{Durech10}, 
none of the oblate objects have an axis ratio $c/a \le0.4$ while 
some prolate ones have an axis ratio $c/a \sim 0.2$. 
This distribution implies an extremely 
elongated ``cigar-like'' shape is more likely than 
an extremely flat ``pancake-like'' shape.  The prolate shape is 
also energetically more stable and permits a much larger 
range of orientations of rotation pole on the sky\cite{Belton18}.

\subsection{Damping timescales.}
The stress-strain cycling caused by the non-principal-axis rotation within 
a non-rigid body will ultimately bring the axis of rotation into alignment 
with its principal axis of the maximum moment of inertia. 
The characteristic damping timescale of the tumbling rotation can be 
estimated by\cite{Harris94}

$\tau_d\simeq\dfrac{P^3}{C_d^3 D^2}$, 

\noindent 
where $P$ and $D$ are the rotation period and mean diameter of the body, 
respectively, and $C_d$ is a constant related to the rigidity, energy 
dissipation property and shape of the body. For $P$ measured in hours, 
$D$ measured in kilometres, and $\tau_d$ measured in Gyr ($10^9$ years), 
the value of $C_d$ is about $17\pm2.5$. For a hundred-meter-sized 
fragment formed by tidal disruption, the rotation period $P$ usually 
ranges from $\sim$5 hours to $\sim$50 hours (see Fig. \ref{Figure2}) 
and the mean diameter $D\simeq2\sqrt[3]{abc}\simeq60$ m, predicting 
$5\times10^9\le\tau_d\le1\times10^{13}$ years, consistent with 
the damping timescales estimated for `Oumuamua\cite{Fraser18}.

\subsection{Thermal modelling.}
To evaluate the temperature evolution of an object 
during close encounters with its original main-sequence 
host star, we use a finite-difference numerical 
technique to model the heat conduction process. 
For comparison with previous analyses for 
`Oumuamua\cite{Fitzsimmons18}, 
the same setup is applied for the thermal modelling, 
in which the object has a comet-like Bond albedo of 
0.01, a bolometric emissivity of 0.95, a thermal
conductivity of 0.001 W m$^{-1}$ K$^{-1}$, a bulk 
density of 1,000 kg m$^{-3}$, and a heat capacity of 
550 J kg$^{-1}$ K$^{-1}$.  Assuming an encounter with 
a star with a mass of $0.5M_{\odot}$, the stellar flux at 
1 AU is about $(0.5)^{3.5}F_{\odot}({\rm1~AU})\simeq121$ W 
m$^{-2}$, where $F_{\odot}$ is the local integrated flux 
of the Sun. The thermal modelling results can be scaled 
to different thermal property values\cite{Fitzsimmons18}. 
To show that volatiles in the subsurface can be 
preserved even under the most extreme condition, we 
intentionally looked at the most intensive illumination
condition. The simulated surface element is assumed to 
be permanently illuminated by the star during the 
encounter, and an orbit with a small periastron 
distance $d_p = 3.5 \times 10^8$ m is used for the 
thermal analyses. 

The simulation was started 153,000 days before 
periastron when the object was over 250 AU away from 
the star, where it had a temperature close to 
the interstellar temperature $\sim$10 K. As shown in 
Fig.~\ref{Figure3}a, the surface temperature 
significantly increases 200 days before periastron, 
but the subsurface region is almost impervious to the 
heat of the star. H$_2$O ice buried $\gtrsim$0.2 m and
CO$_2$ ice buried $\gtrsim$0.5 m beneath the surface 
can be preserved. The thermal simulation tracked the 
temperature evolution until 110,000 days after 
periastron (well beyond the range showed in Fig.~\ref{Figure3}). 
The results show that the slow inward 
propagation of periastron heat can activate the 
sublimation of CO ice buried below a depth of 3 m 
which is a large fraction of `Oumuamua's minor axis $c$. 
The thorough clearing of low-sublimation-temperature 
volatiles accounts for the lack of visible cometary 
activities around `Oumuamua during its subsequent 
perihelion passage through the Solar System in 2017. 

If `Oumuamua is one of the fragments that originated 
from the parent body's undifferentiated interior, most 
of its surface area would be exposed to the star only 
after the periastron passage.  Less heat would be 
transferred to the interior (see Fig.~\ref{Figure3}b)
and more volatiles could be preserved at a shallower 
depth in its interior (H$_2$O ice buried $\gtrsim$0.1
m and CO$_2$ ice buried $\gtrsim$0.2 m beneath the 
surface can be preserved).

Similar models can be constructed for main-sequence 
stars with different mass.  
The flux near the stellar surface is $\propto
M_s^{1.5}$ (because the luminosity is $\propto
M_s^{3.5}$ and $R_s \propto M_s$). 
Grazing passage in the proximity of the stellar surface 
(over a timescale $\propto M_s$) can heat the surfaces 
of intruding objects well above the sublimation temperature 
of the volatile ices.  However, the sub-surface volatile 
deposits may be less efficiently depleted during the 
brief encounters with less luminous low-mass stars.
Around main-sequence hosts, energy dissipation due to 
the internal friction between particles is too weak to 
cause a significant amount of volatile losses.  

 We also consider the possibility of 
white dwarfs as `Oumuamua's original host stars.  
Prior to their emergence, their progenitors go through
a red giant phase when their envelopes engulf 
planets and asteroids within a few AU and their 
radiation increases the equilibrium temperature 
within $\sim$100 AU above most volatiles' 
sublimation temperature. During their subsequent mass 
loss to surrounding planetary nebula, a fraction 
of their outer planets and residual planetesimals remain
attached, though with more expanded orbits, to  
the white dwarfs' gravity\cite{Veras16}. They join the 
reservoir of potential parent bodies of asteroidal ISOs
(see section ``Asteroidal ISOs' mass budget'' in Methods).  

After the planetary nebula is dispersed, the 
luminosity of white dwarfs decreases from hundredths of the 
luminosity in their main-sequence stage at age $\lesssim 0.1$ 
Gyr to ten thousandth of that after 10 Gyr. For these ages, 
the equilibrium temperatures at $d_{\rm limit} (R_p=100
~{\rm m}) \sim 0.5 R_\odot$ are 2600 K and 820 K for a 
$0.5M_\odot$ white dwarf, respectively, 
well above the sublimation temperatures of volatiles. 
The fragments' silicate surfaces may become molten and 
their cohesive strength may be enhanced by the subsequent 
solidification.
For grazing encounters with $d_p \sim R_s$, although 
smaller ($R_p \gtrsim 0.1$ m) objects can withstand the
intense tidal stress, they are expected to be totally 
sublimated with 7 times higher equilibrium temperatures
and accreted onto the white dwarfs in a gas phase\cite{Chen19}. 

During `Oumuamua's recent excursion in the Solar System,
its surface was once again exposed by the Sun's intense 
radiation.
Heat diffusion can activate the sublimation of residual 
H$_2$O and CO$_2$ ices\cite{Weissman83, Micheli18}. 
We simulated the heat conduction process for `Oumuamua
during its Solar System passage using the same thermal 
properties introduced in the first paragraph of this section.
The simulation was started 51,000 days before perihelion 
when `Oumuamua was over 780 AU away from the Sun and 
had a initial temperature of 10 K.
The thermal analyses indicate that, during the span of 
observation, the sublimation temperature of H$_2$O ices 
is reached at $\sim$0.4 m beneath the surface, and that of 
CO$_2$ is reached at $\sim$0.5 m (see Fig.~\ref{Figure3}c), 
consistent with previous results\cite{Fitzsimmons18, Micheli18}.
Due to the consumption of H$_2$O and CO$_2$ during our 
hypothetical formation process, the CO$_2$ buried in subsurface 
may not be activated during the observation,
while the volatile inventory of H$_2$O is sufficient to provide 
the observed non-gravitational acceleration (see section 
``Vaporising H$_2$O ice by solar irradiation to provide the 
non-gravitational acceleration'' in Methods). 
Because of the time delay involved in thermal processes,
the outgassing of H$_2$O in the subsurface should commence 
several days later after reaching sublimation temperature. 
This outgassing activity is consistent with the low CO$_2$ 
production rate during the observation\cite{Trilling18} 
and the non detection of H$_2$O during `Oumuamua's 
perihelion passage\cite{Hui19}.

\subsection{Motion relative to the Local Standard of Rest (LSR).}
Prior to `Oumuamua's passage through the Solar System, its velocity 
relative to the LSR ($v_O \sim 10$ km s$^{-1}$) 
is small compared with that of mature solar type stars\cite{Meech17}. 
Stars in the solar neighbourhood are formed in
turbulent giant molecular clouds (GMCs) that
also gravitationally stir stars to gain velocity
dispersion relatively to the LSR\cite{Spitzer53}.
Their velocity dispersion relative to the LSR is
observed\cite{Holmberg07} to be

$v_d (\tau_s) \sim v_{d0} (\tau_s/\tau_{d0})^{1/2}$,

\noindent
where $v_{d0} \simeq 10$ km s$^{-1}$, $\tau_{d
0} \simeq 1$ Gyr, and $\tau_s$ is the stellar age.
Relative to $v_d$, the modest value of $v_O$ may 
imply `Oumuamua comes from a young stellar 
system, although the possibility that it has been 
travelling in the interstellar medium (ISM) for billions 
of years cannot be ruled out\cite{Meech17,Portegies18}. 
Here we analyse the potential effects of the ISM drag and the coupling 
of motion with magnetic fields on the ISOs' velocity relative 
to the LSR.

The energy per unit mass of ISOs' random motion 
increases at a rate ${\dot e}_{\rm gain} =
(1/2) d v_d^2/d\tau \sim v_{d0}^2/2 \tau_{d0}$.
The motion of ISOs through the ISM 
(with a density $n_{\rm ISM}$) also leads 
to energy dissipation at a rate

${\dot e}_{\rm diss} \sim \dfrac{C_D n_{\rm ISM} m_p 
v_d^3 \pi a c}  {\pi ac^2 \rho_O},$

\noindent
where $C_D$ is the drag coefficient of order 
unity, $m_p$ is the mass of a proton, $c$ and $a$
are the sizes of a prolate ellipsoid with 
semi-minor and semi-major axes (in the range 
of 10--100 m for `Oumuamua), $\rho_O$ is the bulk 
density (assuming $\sim$1,000 kg m$^{-3}$ for `Oumuamua), 
and mass $M_O \sim  \pi \rho_O a c^2$, respectively.  
They reach a terminal speed with 

$v_{\rm term} 
\simeq 
\left( 
\dfrac{ \rho_O c {\dot e}_{\rm gain}}  {C_D 
n_{\rm ISM} m_p}
\right)^{1/3} \sim v_{d0} \left( 
\dfrac{c}{10 {\rm~m}} \dfrac{\rho_O} 
{\rm{10^3}~{\rm kg~m}^{-3}} \dfrac{10^{10}~{\rm m}^{-3}}
{n_{\rm ISM}} \right)^{1/3}$.

Among the ISM's multiple components\cite{Draine11},
the warm ionised or neutral hydrogen gas has
$n_{\rm ISM} \sim 10^6$ m$^{-3}$, which is too
tenuous to make a difference. Although the GMCs have 
$n_{\rm ISM} \sim 10^{10}$--$10^{12}$ m$^{-3}$, 
they occupy $< 1 \%$ of the volume and
cannot be effective decelerators unless `Oumuamua
was trapped in them.  The cold neutral gas has
$n_{\rm ISM} \lesssim 10^8$ m$^{-3}$ and fills a
much larger fractional volume than the GMCs, 
especially near the mid-galactic plane. It can
only reduce $v_{\rm term}$ to be comparable to
that of `Oumuamua's motion\cite{Bialy18,
Eubanks19} if `Oumuamua's $\rho_O \ll1$,000 kg m$^{-3}$.

The ISM also contains magnetic fields with 
$B \sim 10~\mu$G = 1 nT in the neutral medium
and a few times larger in the molecular clouds\cite{Heiles05, Draine11}.
As `Oumuamua moves through the field, 
an electric field ${{\bf v} \times {\bf B}}$ is induced across
it with an electric potential $U \sim a vB$.
An Alfv\'en wave is emitted along a wing\cite{Drell65} with a power  

$P \simeq U^2/R_{\rm wing}= a^2 v^2 B^2/\mu_0 v_A.$

\noindent
In SI units, $R_{\rm wing} \sim \mu_0 v_A$ 
is the effective resistance\cite{Neubauer80}
provided by the surrounding plasma, $\mu_0= 4 \pi \times
10^{-7}$ N A$^{-2}$ is the permeability of free
space, and $v_A 
\sim 1$ km s$^{-1}$ is the Alfv\'en speed. This
power dissipates `Oumuamua's kinetic energy
per unit mass at a rate $d v^2/dt \simeq P/M_O$.
When the power dissipated in the Alfv\'en wing is
balanced by ${\dot e}_{\rm gain}$, a terminal 
speed is established with

\noindent
$\dfrac{v_{\rm term}} {v_{d0}}
\simeq \left( \dfrac{\pi c^2}{2 a} \dfrac{\rho_O}{\tau_{d 0}} 
\dfrac{\mu_0 v_A}{B^2} \right)^{1/2}
\simeq 7 \left(\dfrac{\rho_O}{\rm{10^3}~{\rm kg~m^{-3}}}
\right)^{1/2} \left( \dfrac{c}{10~{\rm m}}
\right)^{1/2} \left( \dfrac{10 c}{a} \right)^{1/2} 
\left( \dfrac{v_A}{\rm 1 \ km \ s^{-1}} \right)
^{1/2} \left( \dfrac{ 1~{\rm nT}}{B} \right)$

\noindent
in SI units. In regions with $B \sim$ a few nT,
small ($c \lesssim 10$ m) ISOs $v_{\rm term}$
is marginally comparable to $v_{d 0} \sim 10$ km
s$^{-1}$ and `Oumuamua's observed velocity
relative to the LSR.  The 
terminal speed for ISOs with $a \sim c \geq 1$
km, is comparable to the dispersion speed with
respect to the LSR for stars with comparable age.
 
In the above estimate, we have assumed the magnetic diffusion 
timescale $\tau_m \sim a^2/ \eta$ is shorter than the tumbling 
timescale $T_\mathrm{tum}$.  
The magnetic diffusivity $\eta$ is $1/\mu_0 \sigma_e$, where 
the electric conductivity $\sigma_e$ of cold rock\cite{Wang13} 
can be as low as $10^{-6} $ S~m$^{-1}$.  In this limit, 
$\tau_m \ll T_\mathrm{tum}$ such that the field is decoupled 
from `Oumuamua's tumbling motion\cite{Laine08}. 
In order to determine the drag on the `Oumuamua's spin (with 
a frequency $\omega_O$) by the Alfv\'en wave\cite{Rafikov99}, 
we consider that its centre of mass is at rest 
with the interstellar magnetic field and its spin axis
is along its minor axis and is inclined to the field
lines at an angle $\theta$.  The spin of `Oumuamua
in the field induces an effective potential drop 
$\sim {\rm sin} \theta B \omega_O a^2/8$ between 
each end and the centre of mass of the spheroid
(over a length $a/2$).  The total power emitted by 
the Alfv\'en waves at both ends of the spheroid
is $\sim {\rm sin}^2 \theta B^2 \omega_O^2 a^4 
/32 \mu_0 v_A$. This loss leads to a change in
`Oumuamua's spin energy $\sim \pi \rho_o
c^2 \omega_O^2 a^3/24$ such that its spin evolves
on a timescale $\tau_\omega={\dot \omega}_O 
/\omega_O \sim (16 / 3 {\rm sin}^2 \theta) 
(v_{\rm term}/ v_{\rm d0})^2 \tau_{d 0}$, which
is greater than a few Gyr.  ISOs' tumbling motion 
through magnetic fields (coupled to the turbulent
interstellar medium) further lengthens $\tau_\omega$,
so that the effect of Alfv\'en wave radiation does 
not significantly affect `Oumuamua's spin frequency.

\subsection{Vaporising H$_2$O ice by solar irradiation to provide 
the non-gravitational acceleration.}
Since other scenarios either require unrealistic 
physical properties or do not lead to sufficient
thrust, outgassing is the most promising mechanism 
to account for `Oumuamua's observed non-gravitational 
acceleration during its brief passage through the 
inner Solar System\cite{Micheli18}.  
Moreover, the light curve of `Oumuamua can also be well 
reproduced with the consideration of the 
torque caused by outgassing\cite{Mashchenko19}.
Upper limits put on the production rate of several 
molecular species of `Oumuamua rule 
out\cite{Ye17, Trilling18, Meech17, Jewitt17} that 
the observed non-gravitational acceleration is 
driven by CN/C$_2$/C$_3$, CO$_2$/CO, and 
micron-size dust.  Nevertheless, H$_2$O remains 
a viable source of outgassing and may potentially account for 
the non-gravitational forces\cite{Micheli18, Park18}.  
For an object with a mass of $m$, the acceleration 
generated from vaporising water ice is 
$a_{\rm H_2O} = Q_{\rm H_2 O} \eta v_{\rm th} / m$, 
where $Q_{\rm H_2 O}$ is the production rate of H$_2$O and 
$v_{\rm th}$ is the gas thermal speed.  
The dimensionless efficiency factor $\eta$ is largely 
affected by the geometry of the emission.  
With a porous surface\cite{Poppe03}, the outgassing of 
H$_2$O molecules into vacuum is expected to be collimated 
towards the Sun\cite{Seligman19} rather than 
in some random directions. 
The effective irradiated area is taken as the projected 
area that is normal to the solar irradiation direction such 
that $\eta \sim 1$.  
During the span of observation (from 2017 October 19 to 2018 
January 2), the surface temperature of `Oumuamua varies from 
350 K to 200 K with $v_{\rm th} \simeq 300$--$400$ m/s. 
The observed non-gravitational radial acceleration of 
`Oumuamua is about 
$a_{\rm ng} \sim 5 \times 10^{-6} (r/1~\rm AU)^{-2}$ m s$^{-2}$. 
For a 200 m $\times$ 20 m $\times$ 20 m ($a=100$ m
and $b=c=10$ m) prolate object with a bulk density of 
1,000 kg m$^{-3}$, the required water sublimation production
rate is $Q_{\rm H_2 O} \simeq m a_{\rm ng} /v_{\rm th}\sim0.7$ 
kg/s at 1 AU.

Assuming the effective irradiated area for a tumbling prolate object 
is $A_{\rm eff}$, which is the averaged projected area 
normal to the solar irradiation direction, the vapour production rate 
can be estimated by

$Q_{\rm H_2 O} ^\prime = \left[
    \dfrac{(1-p) F_\odot - \epsilon \sigma T_{\rm sub}^4}
    {\Delta H / N_A + \gamma k_{B} T_{\rm sub}} \right] A_{\rm eff} m_{\rm H_2 O}$, 

\noindent where $p$ is the surface albedo, $F_\odot$ is the local 
integrated solar flux, $\epsilon$ is the bolometric emissivity, 
$\sigma$ is the Stefan-Boltzmann constant, $T_{\rm sub}$ is the 
sublimation temperature, $\Delta H$ is the sublimation enthalpy, 
$N_A$ is the Avogadro constant, $\gamma$ is the heat capacity ratio, 
$k_B$ is the Boltzmann constant, and $m_{\rm H_2O}$ is the mass of 
one H$_2$O molecule. 
With $p=0$, $F_\odot = 1367 (r/{\rm 1 AU})^{-2}$ W m$^{-2}$, 
$\epsilon =0.95$, $T_{\rm sub}= 144$ K, $\Delta H=54000$ J/mol, 
$\gamma=1.33$, and $A_{\rm eff} = \pi a b$, the maximum production 
rate at 1AU is $\sim 1.37$ kg/s. This value is comparable to 
the required water vapour production rate $Q_{\rm H_2 O}$
to account for the non-gravitational acceleration of `Oumuamua. 

During the two months' observations since `Oumuamua's perihelion 
passage, the total outgassing amount of H$_2$O to produce the 
non-gravitational acceleration is $\sim$$1.2\times10^6$ kg. 
The water ice inventory between $\sim$0.1--0.2 m and $\sim$0.4 m that 
can commence sublimation during this period (see Fig.~\ref{Figure3}c), 
is sufficient to provide the required amount of H$_2$O.
This amount is a few percent of $M_O$ so that `Oumuamua's mass and size 
were not significantly modified.

Nevertheless, a low production rate of water ice has been suggested for
`Oumuamua based on its low observationally inferred carbon abundance and 
the C/O ratios of some Solar System comets\cite{Sekanina19}.  
If most of the volatile carbon-rich molecules have already sublimated 
and been ejected from `Oumuamua when it was in the proximity of its 
original host star (see section ``Thermal modelling'' in Methods), 
this constraint would be relaxed.  Observations also indicate low C/O 
ratios in extrasolar planetesimals\cite{Wilson16}.

With a porous surface, dust production associated with the gas sublimation 
can also lead to outgassing with silicate-water ice composition.  This 
additional source would enhance `Oumuamua's non-gravitational acceleration. 
If the grains are predominantly larger than a few hundred micrometres to 
millimetres, they would not have been detected at optical 
wavelengths\cite{Meech17, Jewitt17, Ye17, Micheli18}. 
Other volatile gases including N$_2$ and H$_2$ can sublimate at lower 
temperatures with lower sublimation enthalpy. They are also more likely 
to be severely depleted during the debris' close encounter with their 
original host stars.  Such propulsion fuel may need to be accreted 
onto ISOs during their voyage through the GMCs\cite{Sandford93}. 

\subsection{Asteroidal ISOs' mass budget.}
Based on the sole detection of `Oumuamua during 3.5 yr
operation of the Panoramic Survey Telescope and Rapid Response 
System (Pan-STARRS),  the spacial density of asteroidal ISOs with 
`Oumuamua's attributes is estimated\cite{Do18, Portegies18} to be 
$n_{\rm ISO} \sim 3.5\times 10^{13}-2 \times 10^{15}$ pc$^{-3}$. 
The range reflects uncertainties on detection efficiency and Sun's
gravitational focusing effect. 

There are potential candidates for ISOs' original host stars in 
the solar neighbourhood. The average stellar mass\cite{Chabrier03} 
and number density\cite{Portegies18} are $M_\ast \sim 0.37 M_\odot$ 
and $n_\ast \sim 0.32$ pc$^{-3}$, respectively. 
In order to attain $n_{\rm ISO}$, each star must eject, on average, 
$N_{\rm ej} \simeq n_{\rm ISO}/n_\ast \sim (1$--$60) \times 10^{14}$ 
asteroidal ISOs.  
Using `Oumuamua's mass ($M_O \lesssim 10^8$ kg for $\rho_O = 1$,000 
kg m$^{-3}$, $a=100$ m, $b=c=10$ m) to normalise the ISOs' average 
mass $M_\mathrm{ISO}$, the required mass ejection per star is 
$M_{\rm ej} = N_{\rm ej} M_{\rm ISO} \simeq (0.0017$--$0.1) (M_{\rm ISO}/M_O) M_\oplus$,
i.e., a fraction of the mass of the Moon or the Earth. Even for
the upper limit, our estimate is two orders of magnitude 
less than some previous estimates\cite{Do18} because we 
use a smaller ISO mass normalisation factor $M_O$ (than
that of a sphere with $a=b=c=100$ m and $\rho_O=3$,000 
kg m$^{-3}$) and a larger $n_\ast$ (relative
to $0.1$ pc$^{-3}$). The magnitude of $M_{\rm ej}$ would 
be further reduced if `Oumuamua's small speed relative to
the LSR (in comparison to the old field stars) leads to a
local concentration of sub-km-sized ISOs.

We now compare $M_{\rm ej}$ with various potential
reservoirs of their progenitors. {\it In situ} 
Solar System formation scenario\cite{Hayashi85} 
suggests a minimum mass $\sim$$100 M_\oplus$ of 
planetesimals within a few tens of AU.  Similar 
mass values are estimated for protostellar disks.  
Since stars form in clusters with neighbours which perturb
the outer ($\gtrsim 100$--1,000 AU) disk regions,  fractional 
loss of residual planetary building blocks is generally 
expected\cite{Portegies18}. However, the detached planetesimals 
from these regions are mostly icy, similar to the long-period 
comets (LPCs).  Without surface transformation due
to heating during their close encounters with their host 
stars, the ejected objects become cometary ISOs,
which may be the mechanism to eject the interstellar 
comet 2I/Borisov\cite{Guzik19} from its original planetary system.

During advanced stages of their growth, proto gas
and ice giant planets scatter\cite{Zhoulin07} 
a few $M_\oplus$ of nearby residual planetesimals. 
A fraction of the outwardly scattered objects settle in 
the inner and outer Oort cloud (with orbital semi-major 
axes $\sim$$2\times10^3$--$2\times10^4$ AU and 
$\sim$$2\times10^4$--$2\times10^5$ AU, respectively) as LPCs. 
From the average mass\cite{Weissman97} 
($M_{\rm comet} \sim 4 \times 10^{13}$ kg) of 
individual LPCs and their estimated total 
number\cite{Kaib09} ($N_{\rm LPC} \sim 10^{12}$), 
we infer a total mass $M_{\rm LPC}= N_{\rm LPC} 
M_{\rm comet} \sim 7 M_\oplus$ for the LPC population.

In comparison with required asteroidal ISOs to be ejected 
per star, $N_{\rm LPC} \sim (0.00017$--$0.01) N_{\rm ej}$, 
$M_{\rm LPC} \gtrsim (70$--$\mathrm{4,000}) M_{\rm ej}$, 
because $M_{\rm comet} \gtrsim 4 
\times 10^5 M_O$. In order for the LPCs to evolve into 
sufficient number of `Oumuamua-like ISOs, they need to
be efficiently scattered to the proximity of the host 
star and substantially downsized analogous to comet 
Shoemaker-Levy-9.  LPCs in the outer Oort cloud are 
loosely bound to the Sun whereas those stored in the 
inner Oort cloud are weakly perturbed by the combination 
of the Galactic tide, passing stars, gas and ice giant 
planets. These effects can lead to the loss of half of 
LPC population to the interstellar space over the Sun's
main sequence lifespan\cite{Hanse17}.  They also lead 
to an infusion of LPCs to the Sun's proximity\cite{Kaib09}
at the observed rate of $\sim1$--10 
events per year\cite{Everhart67, Francis05, Neslusan07}.

Similar dynamical processes can also occur around 
lower-mass stars. Although they rarely bear gas giants,
perturbations from the Galactic tides, field stars,
ice giant planets, and super-Earths may have more intense
influence on the dynamical evolution of residual 
planetesimals around lower-mass stars. Since $M_{\rm LPC}
\sim 100 M_{\rm ej}$, they have adequate supply of
potential parent bodies for producing asteroidal ISOs 
by our hypothetical formation and ejection mechanism, even in 
the limit that the mass of their debris disk is 
proportion to their own mass.

As the host stars evolve off the main-sequence stage,
the red-giant envelopes engulf planets and residual
planetesimals within a few AU.  While most of the
outer major, dwarf, and minor planets are released 
to become freely floating planets and cometary ISOs,
a fraction may be retained\cite{Veras11}.  The mass
reduction of their hosts also promotes long-term orbital 
evolution\cite{Raymond18a, Rafikov18}, which may lead to 
close stellar encounters and tidal disruption. 
Many white dwarfs show signs of ongoing and active 
accretion of refractory elements\cite{Manser19, Chen19} 
at the rate of $\sim$$10^5$--$10^6$ kg s$^{-1}$. 
Over the white dwarfs' cooling and 
dimming timescales, the integrated amount of mostly
refractory elements accreted onto them\cite{Chen19} 
is $\sim$0.001--0.1$M_\oplus$.  If a similar amount
of $100$-m-sized tidal disrupted fragments are released
from their host white dwarf, it would be comparable to 
the lower estimates for $M_{\rm ej}$. However, parent 
bodies of asteroidal ISOs in this size range are 
preserved provided their $d_p \sim d_{\rm limit} 
(100~{\rm m}) \sim 0.5 R_\odot$ (see ``Tidal disruption 
limiting distance'' in Methods).  
This requirement confines the formation domain around white
dwarfs for the `Oumuamua-like asteroidal ISOs.  Finally,
the progenitors of white dwarfs are main-sequence stars 
with mass in the range (0.8--8)$M_\odot$.  They are an 
order of magnitude less numerous than the lower-mass 
main-sequence stars.  Nevertheless, they may provide the
original host for a small fraction of the asteroidal ISOs
with `Oumuamua's attributes.

\subsection{Tidal disintegration of super-Earths 
around low-mass stars.}
In the main text, we introduce 
the tidal disruption simulations of a self-gravity-dominated 
super-Earth (with $R_p = 2 \times 10^7$ m and $M_p = 11 
M_\oplus$) on an $e_0 = 0.999$ and $d_p = 4 \times 10^8$ m
$< d_{\rm Roche}$ orbit around a 0.5$M_\odot$ host star.
In this case, half of the fragment, with mass up to
$M_{\rm upper} \sim 0.1 M_p$, can obtain a normalised
orbital-energy-per-unit-mass increment $f_E \gtrsim 1$
and escape from the host star (see ``Ejection condition'' 
in Methods). 
The ejection of such large fragments provides a potential 
release mechanism for the freely floating super-Earths found  
by the microlensing observations\cite{Mroz19}. 
The retained fragments with eccentricity $<1$ and mass 
$\lesssim M_{\rm upper}$, come back for subsequent 
encounters and be ground down to smaller fragments.
These later-generation fragments also 
contribute to the population of asteroidal ISOs.  But, 
if multiple downsizing steps are needed, only a minute 
fraction of the initial $M_p$ would be transformed into 
sub-km-sized asteroidal ISOs.  

In Methods section ``Tidal disruption limiting distance'', 
we show that the condition for tidal disruption
is $R_p$-independent for large objects bound by gravity.
But for small strength-dominated bodies, the 
magnitude of $d_{\rm limit} (R_p)$ is an increasing 
function of $R_p$.  In the limit that $d_{\rm Roche} 
\gtrsim d_p \gtrsim R_s$, the disintegration of 
planet-size objects continues to cascade until their 
fragments are reduced to the size $R_p$ for which 
$d_{\rm limit} (R_p) \sim R_s$.  Smaller fragments 
preserve their integrity outside the stellar surface.
Since $d_{\rm limit}/R_s
\propto M^{-2/3}$, the region where tidal disruption 
occurs is more extended around low-$M_s$ stars.

We simulate the consequence of more intense tidal
disruption with $d_p=1.0, 1.2,$ and $1.5 \times 10^8$ m 
around a $M_s=0.1M_\odot$ and $R_s \sim 0.1 R_\odot$ 
main-sequence star whose $d_{\rm Roche} \sim 4 \times 
10^8$ m. Our results show that, for $d_p = 1.5 \times 
10^8$ m, the largest fragments' mass $M_{\rm upper}
\sim 0.1 M_p$. 
But, with $d_p = 1.0\times 10^8$ m, the parent body breaks
down to the numerical resolution limit with
$M_{\rm upper} \lesssim 6 \times 10^{-3} M_p$
(see Supplementary Fig.~6).
The size distribution of the fragments almost matches
its initial particles' size distribution.
In order to overcome the indicative limitation on numerical
resolution, we simulated tidal disruption of
smaller (with $R_p= 10^5$ m, $10^4$ m, and $10^3$ m) parent 
bodies. When the cohesive material strength is
neglected, the simulations give very similar results to 
the $2 \times 10^7$ m model (these simulations do include
the shear strength, which is size-independent).

This powderization process ceases when the size of 
fragments enters the strength-dominated regime\cite{Guimaraes12}. 
For illustration, we impose cohesion in simulations with 
$C = 10$ Pa using the same encounter condition. In this 
limit, the model with $R_p=$ 1 km produces larger (a few 
hundred meter-size) fragments than the cohesion-free 
model.  The required cohesive strength for preserving the
integrity of a fragment can be estimated by

$C \ge \dfrac{4\pi G R_p^2\rho_p^2}{5} \bigg( 
\dfrac{\sqrt{3}M_s}{4\pi \rho_s d_p^3} - s \bigg).$

\noindent
Hundred-meter-sized fragments with
$C\sim50$ Pa can survive in this extreme strong
tidal disruption event.  This level of cohesive
strength can be easily provided by the sintering
bonds developed during the surface solidification 
shortly after the periastron passage\cite{Schwartz13}. 
By this means, most of the escaped fragments should have a 
size in the range of dozen meters to hundred meters, one of 
which could be an `Oumuamua analog.

We estimate the probability of producing asteroidal 
ISOs by the tidal downsizing of dwarf planets, 
super-Earths and ice giants. 
Most low-mass stars bear multiple super-Earths 
and sub-Neptunes with mass on the order of a few $M_\oplus$ 
($\sim$$100 M_{\rm ej}$).  The prospect of tapping these 
rich reservoirs to match $M_{\rm ej}$ requires a few 
percent efficiency for the delivery of a single 
super-Earths (or high efficiency of delivering a single dwarf 
planet) to the proximity of each individual host star on a 
sufficiently high-eccentricity orbit.  
Some potentially efficient triggering mechanisms 
include close encounters with, secular or Lidov-Kosai 
resonances induced by their binary stellar and gas 
giant planetary companions or by perturbations from passing 
field stars\cite{Nagasawa08, Wu11, Punzo14, Portegies18}.

Analogous to the Earth, Uranus, and Neptune, the molten 
cores of exoplanets probably contain little CO but a 
vast amount of water under some super-critical phases. 
Planetary fragments' water retention probability and 
outgassing efficiency under the condition of their 
formation and subsequent intrusion into the Solar System
are poorly known.  It is highly uncertain whether the 
mechanism we proposed for `Oumuamua's non-gravitational 
acceleration can operate among the relics of 
molten planetary cores.

\end{methods}

\section*{Data availability} The data that support the plots within this paper and other findings of this study are available from the corresponding author upon reasonable request.

\section*{Code availability} The code used for the thermal analyses is available from the corresponding author upon reasonable request.  The PKDGRAV code with granular physics is not yet ready for public release - its details and validation have been presented in many previous studies and are available from the corresponding author upon reasonable request.



\begin{thebibliography}{10}
\bibitem{Bannister19} Bannister, M. T. et al. The natural history of `Oumuamua. \textit{Nature Astron.} \textbf{3}, 594--602 (2019).
\bibitem{Meech17} Meech, K. J. et al. A brief visit from a red and extremely elongated interstellar asteroid. \textit{Nature} \textbf{552}, 378--381 (2017).
\bibitem{Knight17} Knight, M. M. et al. On the rotation period and shape of the hyperbolic asteroid 1I/`Oumuamua (2017 U1) from its lightcurve. \textit{Astrophys. J.} \textbf{851}, L31 (2017).
\bibitem{Bolin17} Bolin, B. T. et al. Apo time-resolved color photometry of highly elongated interstellar object 1I/`Oumuamua. \textit{Astrophys. J.} \textbf{852}, L2 (2017).
\bibitem{Fraser18} Fraser, W. C. et al. The tumbling rotational state of 1I/`Oumuamua. \textit{Nature Astron.} \textbf{2}, 383--386 (2018).
\bibitem{Drahus18} Drahus,  M. et  al. Tumbling  motion  of  1I/`Oumuamua  and  its  implications  for  the  body's distant past. \textit{Nature Astron.} \textbf{2}, 407--412 (2018).
\bibitem{Micheli18} Micheli, M. et al. Non-gravitational acceleration in the trajectory of 1I/2017 U1 (`Oumuamua). \textit{Nature} \textbf{559}, 223--226 (2018).
\bibitem{Trilling18} Trilling, D. E. et al. Spitzer observations of interstellar object 1I/`Oumuamua. \textit{Astron. J.} \textbf{156}, 261 (2018).
\bibitem{Jewitt17} Jewitt, D. et al. Interstellar interloper 1I/2017 U1: observations from the NOT and WIYN telescopes. \textit{Astrophys. J.} \textbf{850}, L36 (2017).
\bibitem{Do18} Do, A., Tucker, M. A. \& Tonry, J. Interstellar interlopers: number density and origin of `Oumuamua-like objects. \textit{Astrophys. J.} \textbf{855}, L10 (2018).
\bibitem{Portegies18} Portegies Zwart, S., Torres, S., Pelupessy, I., B{\'e}dorf, J. \& Cai, M. X. The origin of interstellar asteroidal objects like 1I/2017 U1 `Oumuamua. \textit{Mon. Not. R. Astron. Soc.} \textbf{479}, L17--L22 (2018).
\bibitem{Engelhardt17} Engelhardt, T. et al. An observational upper limit on the interstellar number density of asteroids and comets. \textit{Astron. J.} \textbf{153}, 133 (2017).
\bibitem{Weissman97} Weissman, P. R. \& Levison, H. F. Origin and evolution of the unusual object 1996 PW: asteroids from the Oort cloud? \textit{Astrophys. J.} \textbf{488}, L133--L136 (1997).
\bibitem{Cuk18} {\'C}uk, M. 1I/`Oumuamua as a tidal disruption fragment from a binary star system. \textit{Astrophys. J.} \textbf{852}, L15 (2018).
\bibitem{Raymond18b} Raymond, S. N., Armitage, P. J. \& Veras, D. Interstellar object `Oumuamua as an extinct fragment of an ejected cometary planetesimal. \textit{Astrophys. J.} \textbf{856},  L7(2018).
\bibitem{Richardson02} Richardson, D. C., Leinhardt, Z. M., Melosh, H. J., Bottke, W. F. \& Asphaug, E. in \textit{Asteroids III} (eds Bottke, W. F. et al.) 501--515 (Univ. of Arizona Press, Tucson, 2002).
\bibitem{Mcneill18} McNeill, A., Trilling, D. E. \& Mommert, M.  Constraints on the density and internal strength of 1I/`Oumuamua. \textit{Astrophys. J.} \textbf{857}, L1 (2018).
\bibitem{Sridhar92} Sridhar, S. \& Tremaine, S. Tidal disruption of viscous bodies. \textit{Icarus} \textbf{95}, 86--99 (1992).
\bibitem{Drell65} Drell, S. D., Foley, H. M. \& Ruderman, M. A. Drag and propulsion of large satellites in the ionosphere: an Alfv{\'{e}}n propulsion engine in space. \textit{J. Geophys. Res.} \textbf{70}, 3131--3145 (1965).
\bibitem{Fitzsimmons18} Fitzsimmons, A. et al. Spectroscopy and thermal modelling of the first interstellar object 1I/2017 U1 `Oumuamua. \textit{Nature Astron.} \textbf{2}, 133--137 (2018).
\bibitem{Kaib09} Kaib, N. A. \& Quinn, T. Reassessing the source of long-period comets. \textit{Science} \textbf{325}, 1234--1236 (2009).
\bibitem{Punzo14} Punzo, D., Capuzzo-Dolcetta, R. \& Portegies Zwart, S. The secular evolution of the Kuiper belt after a close stellar encounter. \textit{Mon. Not. R. Astron. Soc.} \textbf{444}, 2808--2819 (2014).
\bibitem{Coughlin16} Coughlin, J. L. et al. Planetary candidates observed by Kepler. VII. The first fully uniform catalog based on the entire 48-month data set (Q1-Q17 DR24). \textit{Astrophys. J. Suppl. Ser.} \textbf{224}, 12 (2016).
\bibitem{Cassan12} Cassan, A. et al. One or more bound planets per milky way star from microlensing observations. \textit{Nature} \textbf{481}, 167--169 (2012).
\bibitem{Nagasawa08} Nagasawa, M., Ida, S. \& Bessho, T. Formation of hot planets by a combination of planet scattering, tidal circularization, and the Kozai mechanism. \textit{Astrophys. J.} \textbf{678}, 498--508 (2008).
\bibitem{Idaetal13} Ida, S., Lin, D. N. C. \& Nagasawa, M. Toward a deterministic model of planetary formation. VII. Eccentricity distribution of gas giants. \textit{Astrophys. J.} \textbf{775}, 42 (2013).
\bibitem{Charbonneau09} Charbonneau, D. et al. A super-Earth transiting a nearby low-mass star. \textit{Nature} \textbf{462}, 891--894 (2009).
\bibitem{Davies14} Davies, M. B. et al. in \textit{Protostars and Planets VI} (eds Beuther, H. et al.) 787--809 (Univ. of Arizona Press, Tucson, 2014).
\bibitem{Rafikov18} Rafikov, R. R. 1I/2017 `Oumuamua-like interstellar asteroids as possible messengers from dead stars. \textit{Astrophys. J.} \textbf{861}, 35 (2018).
\bibitem{Veras11} Veras, D., Wyatt, M. C., Mustill, A. J., Bonsor, A. \& Eldridge, J. J. The great escape: how exoplanets and smaller bodies desert dying stars. \textit{Mon. Not. R. Astron. Soc.} \textbf{417}, 2104--2123 (2011).
\end{thebibliography}

\begin{thebibliography}{10}
\bibitem[31]{Gasc17} Gasc, S. et al. Change of outgassing pattern of 67P/Churyumov-Gerasimenko during the March 2016 equinox as seen by ROSINA. \textit{Mon. Not. R. Astron. Soc.} \textbf{469}, S108--S117 (2017).
\bibitem[32]{Lesher15} Lesher, C. E. \& Spera, F. J. in \textit{The Encyclopedia of Volcanoes} (eds Sigurdsson, H. et al.) 113--141 (Academic Press, Elsevier, Amsterdam, 2015).
\bibitem[33]{Richardson00} Richardson, D. C., Quinn, T., Stadel, J. \& Lake, G. Direct large-scale \textit{N}-body simulations of planetesimal dynamics. \textit{Icarus} \textbf{143}, 45--59 (2000). 
\bibitem[34]{Stadel01} Stadel, J. G. \textit{Cosmological N-body Simulations and Their Analysis}. Ph.D. thesis, Univ. Washington (2001).
\bibitem[35]{Schwartz12} Schwartz, S. R., Richardson, D. C. \& Michel, P. An implementation of the soft-sphere discrete element method in a high-performance parallel gravity tree-code. \textit{Granul. Matter} \textbf{14}, 363--380 (2012).
\bibitem[36]{Zhang17} Zhang, Y. et al. Creep stability of the proposed AIDA mission target 65803 Didymos: I. discrete cohesionless granular physics model. \textit{Icarus} \textbf{294}, 98--123 (2017).
\bibitem[37]{Zhang18} Zhang, Y. et al. Rotational failure of rubble-pile bodies: influences of shear and cohesive strengths. \textit{Astrophys. J.} \textbf{857}, 15 (2018).
\bibitem[38]{Chau02} Chau, K. T., Wong, R. H. C. \& Wu, J. J. Coefficient of restitution and rotational motions of rockfall impacts. \textit{Int. J. Rock Mech. Min. Sci.} \textbf{39}, 69--77 (2002).
\bibitem[39]{Jiang15} Jiang, M., Shen, Z. \& Wang, J. A novel three-dimensional contact model for granulates incorporating rolling and twisting resistances. \textit{Comput. Geotech.} \textbf{65}, 147--163 (2015).
\bibitem[40]{Schwartz13} Schwartz, S. R., Michel, P. \& Richardson, D. C. Numerically simulating impact disruptions of cohesive glass bead agglomerates using the soft-sphere discrete element method. \textit{Icarus} \textbf{226}, 67--76 (2013).
\bibitem[41]{Poppe03} Poppe, T. Sintering of highly porous silica-particle samples: analogues of early Solar-System aggregates. \textit{Icarus} \textbf{164}, 139--148 (2003).
\bibitem[42]{Holsapple08} Holsapple, K. A. \& Michel, P. Tidal disruptions II: a continuum theory for solid bodies with strength, with applications to the Solar System. \textit{Icarus} \textbf{193}, 283--301 (2008).
\bibitem[43]{Ramirez15} Ram{\'{\i}}rez, I. et al. The dissimilar chemical composition of the planet-hosting stars of the XO-2 binary system. \textit{Astrophys. J.} \textbf{808}, 13 (2015).
\bibitem[44]{Kalirai09} Kalirai, J. S. et al. The masses of population II white dwarfs. \textit{Astrophys. J.} \textbf{705}, 408--425 (2009).
\bibitem[45]{Scheeres00} Scheeres, D. J., Ostro, S. M., Werner, R. A., Asphaug, E. \& Hudson, R. S. Effects of gravitational interactions on asteroid spin states. \textit{Icarus} \textbf{147}, 106--118 (2000).
\bibitem[46]{Belton18} Belton, M. J. et al. The excited spin state of 1I/2017 U1 `Oumuamua. \textit{Astrophys. J.} \textbf{856}, L21 (2018).
\bibitem[47]{Mashchenko19} Mashchenko, S. Modelling the light curve of `Oumuamua: evidence for torque and disc-like shape. \textit{Mon. Not. R. Astron. Soc.} \textbf{489}, 3003-3021 (2019).
\bibitem[48]{Durech10} {\v{D}}urech, J., Sidorin, V. \& Kaasalainen, M. DAMIT: a database of asteroid models. \textit{Astron. Astrophys.} \textbf{513}, A46 (2010).
\bibitem[49]{Harris94} Harris, A. W. Tumbling asteroids. \textit{Icarus} \textbf{107}, 209--211 (1994).
\bibitem[50]{Veras16} Veras, D. Post-main-sequence planetary system evolution. \textit{R. Soc. Open Sci.} \textbf{3}, 150571 (2016).
\bibitem[51]{Chen19} Chen, D.-C. et al. A power-law decay evolution scenario for polluted single white dwarfs. \textit{Nature Astron.} \textbf{3}, 69--75 (2019).
\bibitem[52]{Weissman83} Weissman, P. R. Cometary impacts with the Sun:  physical and dynamical considerations. \textit{Icarus} \textbf{55}, 448--454 (1983).
\bibitem[53]{Hui19} Hui, M.-T., \& Knight, M. M. New insights into interstellar object 1I/2017 U1 (`Oumuamua) from SOHO/STEREO nondetections. \textit{Astron. J.} \textbf{158}, 256 (2019).
\bibitem[54]{Spitzer53} Spitzer, L. \& Schwarzschild, M. The possible influence of interstellar clouds on stellar velocities. II. \textit{Astrophys. J.} \textbf{118}, 106 (1953).
\bibitem[55]{Holmberg07} Holmberg, J., Nordstr{\"o}m, B. \& Andersen, J. The Geneva-Copenhagen survey of the Solar neighbourhood II. New \textit{uvby} calibrations and rediscussion of stellar ages, the G dwarf problem, age-metallicity diagram, and heating mechanisms of the disk. \textit{Astron. Astrophys.} \textbf{475}, 519--537 (2007).
\bibitem[56]{Draine11} Draine, B. T. \textit{Physics of the Interstellar and Intergalactic Medium} (Princeton Univ. Press, 2011).
\bibitem[57]{Bialy18} Bialy, S. \& Loeb, A. Could solar radiation pressure explain `Oumuamua's peculiar acceleration? \textit{Astrophys. J.} \textbf{868}, L1 (2018).
\bibitem[58]{Eubanks19} Eubanks, T. M. High-drag interstellar objects and galactic dynamical streams. \textit{Astrophys. J.} \textbf{874}, L11 (2019).
\bibitem[59]{Heiles05} Heiles, C. \& Crutcher, R. Magnetic fields in diffuse HI and molecular clouds. \textit{Lect. Notes Phys.} \textbf{664}, 137--182 (2005).
\bibitem[60]{Neubauer80} Neubauer, F. M. Nonlinear standing Alfv{\'{e}}n wave current system at Io: theory. \textit{J. Geophys. Res.} \textbf{85}, 1171--1178 (1980).
\bibitem[61]{Wang13} Wang, D. \& Karato, S.-i. Electrical conductivity of talc aggregates at 0.5 GPa: influence of dehydration. \textit{Phys. Chem. Miner. } \textbf{40}, 11--17 (2013).
\bibitem[62]{Laine08} Laine, R. O., Lin, D. N. C. \& Dong, S. Interaction of close-in planets with the magnetosphere of their host stars. I. Diffusion, ohmic dissipation of time-dependent field, planetary inflation, and mass loss. \textit{Astrophys. J.} \textbf{685}, 521--542 (2008).
\bibitem[63]{Rafikov99} Rafikov, R. R., Gurevich, A. V. \& Zybin, K. P. Inductive interaction of rapidly rotating conductive bodies with a magnetized plasma. \textit{J. Exp. Theor. Phys.} \textbf{88}, 297--308 (1999).
\bibitem[64]{Ye17} Ye, Q.-Z., Zhang, Q., Kelley, M. S. P. \& Brown, P. G. 1I/2017 U1 (`Oumuamua) is hot: imaging, spectroscopy, and search of meteor activity. \textit{Astrophys. J.} \textbf{851}, L5 (2017).
\bibitem[65]{Park18} Park, R. S., Pisano, D. J., Lazio, T. J. W., Chodas, P. W. \& Naidu, S. P. Search for OH 18 cm radio emission from 1I/2017 U1 with the Green Bank Telescope. \textit{Astron. J.} \textbf{155}, 185 (2018).
\bibitem[66]{Seligman19} Seligman, D., Laughlin, G. \& Batygin, K. On the anomalous acceleration of 1I/2017 U1 `Oumuamua. \textit{Astrophys. J.} \textbf{876}, L26 (2019).
\bibitem[67]{Sekanina19} Sekanina, Z. Outgassing as trigger of 1I/`Oumuamua's nongravitational acceleration: Could this hypothesis work at all? Preprint at \url{http://arxiv.org/abs/1905.00935} (2019).
\bibitem[68]{Wilson16} Wilson, D. J., G{\"a}nsicke, B. T., Farihi, J. \& Koester, D. Carbon to oxygen ratios in extrasolar planetesimals. \textit{Mon. Not. R. Astron. Soc.} \textbf{459}, 3282--3286 (2016).
\bibitem[69]{Sandford93} Sandford, S. A., Allamandola, L. J. \& Geballe, T. R. Spectroscopic detection of molecular hydrogen frozen in interstellar ices. \textit{Science} \textbf{262}, 400--402 (1993).
\bibitem[70]{Chabrier03} Chabrier, G. Galactic stellar and substellar initial mass function. \textit{Publ. Astron. Soc. Pacif.} \textbf{115}, 763--795 (2003).
\bibitem[71]{Hayashi85} Hayashi, C., Nakazawa, K. \& Nakagawa, Y. in \textit{Protostars and Planets II} (eds Black, D. C. \& Matthews, M. S.) 1100--1153 (1985).
\bibitem[72]{Guzik19} Guzik, P. et al. Initial characterization of interstellar comet 2I/Borisov. \textit{Nature Astron.} \textbf{4}, 53--57 (2019).
\bibitem[73]{Zhoulin07} Zhou, J.-L. \& Lin, D. N. C. Planetesimal accretion onto growing proto-gas giant planets. \textit{Astrophys. J.} \textbf{666}, 447--465 (2007).
\bibitem[74]{Hanse17} Hanse, J., J\'ilkov\'a, L., Portegies Zwart, S. F. \& Pelupessy, F. I. Capture of exocomets and the erosion of the Oort cloud due to stellar encounters in the Galaxy. \textit{Mon. Not. R. Astron. Soc.} \textbf{473}, 5432--5445 (2017).
\bibitem[75]{Everhart67} Everhart, E. Intrinsic distributions of cometary perihelia and magnitudes. \textit{Astron. J.} \textbf{72}, 1002--1011 (1967).
\bibitem[76]{Francis05} Francis, P. J. The demographics of long-period comets. \textit{Astrophys. J.} \textbf{635}, 1348--1361 (2005).
\bibitem[77]{Neslusan07} Neslu{\v{s}}an, L. The fading problem and the population of the Oort cloud. \textit{Astron. Astrophys.} \textbf{461}, 741--750 (2007).
\bibitem[78]{Raymond18a} Raymond, S. N., Armitage, P. J., Veras, D., Quintana, E. V. \& Barclay, T. Implications of the interstellar object 1I/`Oumuamua for planetary dynamics and planetesimal formation. \textit{Mon. Not. R. Astron. Soc.} \textbf{476}, 3031--3038 (2018).
\bibitem[79]{Manser19} Manser, C. J. et al. A planetesimal orbiting within the debris disc around a white dwarf star. \textit{Science} \textbf{364}, 66--69 (2019).
\bibitem[80]{Mroz19} Mr{\'o}z P. et al. Two new free-floating or wide-orbit planets from microlensing. \textit{Astron. Astrophys.} \textbf{622}, A201 (2019).
\bibitem[81]{Guimaraes12} Guimar{\~a}ess, A. H. F. et al. Aggregates in the strength and gravity regime: particles sizes in Saturn's rings. \textit{Icarus} \textbf{220}, 660--678 (2012).
\bibitem[82]{Wu11} Wu, Y. \& Lithwick, Y. Secular chaos and the production of hot Jupiters. \textit{Astrophys. J.} \textbf{735}, 109 (2011).
\end{thebibliography}
\end{document}